\documentclass[conference]{IEEEtran}
\IEEEoverridecommandlockouts
\usepackage{cite}
\usepackage{amsmath,amssymb,amsfonts}
\usepackage{algorithmic}
\usepackage{graphicx}
\usepackage{textcomp}
\usepackage{orcidlink}
\usepackage{xcolor}
\usepackage{array}
\usepackage{tabularx}
\usepackage{multirow}
\usepackage{caption}
\usepackage{booktabs}
\usepackage{hyperref}
\usepackage[T1]{fontenc}
\usepackage{fancyhdr}
\def\BibTeX{{\rm B\kern-.05em{\sc i\kern-.025em b}\kern-.08em
    T\kern-.1667em\lower.7ex\hbox{E}\kern-.125emX}}

\begin{document}

\title{Recovery of Missing Sensor Data by Reconstructing Time-varying Graph Signals \vspace{-1mm}}

\author{{\href{https://sites.google.com/view/anindyamondal}{Anindya Mondal}$^*$\thanks{*Corresponding author: anindyam.jan@gmail.com}, Mayukhmali Das, Aditi Chatterjee, P. Venkateswaran} \\
\textit{\href{http://www.jaduniv.edu.in/view_department.php?deptid=84}{Department of Electronics and Telecommunication Engineering}, \href{http://www.jaduniv.edu.in/}{Jadavpur University}, India} \\
}

\maketitle

\makeatletter
%%%%%%%%%%%%%%%%%%%%%%%%%%%%%% User specified LaTeX commands.
%\def\ps@IEEEtitlepagestyle{%
%  \def\@oddfoot{\mycopyrightnotice}%
%%  \def\@evenfoot{}%
%}
%\def\mycopyrightnotice{%
%  {\footnotesize \textbf{978-1-6654-0962-9/21/\$31.00 \textcopyright 2021 IEEE}\hfill}% <--- Change here
%  \gdef\mycopyrightnotice{}% just in case
%}

\begin{abstract}
Wireless sensor networks are among the most promising technologies of the current era because of their small size, lower cost, and ease of deployment. With the increasing number of wireless sensors, the probability of generating missing data also rises. This incomplete data could lead to disastrous consequences if used for decision-making. There is rich literature dealing with this problem. However, most approaches show performance degradation when a sizable amount of data is lost. Inspired by the emerging field of graph signal processing, this paper performs a new study of a Sobolev reconstruction algorithm in wireless sensor networks. Experimental comparisons on several publicly available datasets demonstrate that the algorithm surpasses multiple state-of-the-art techniques by a maximum margin of 54\%. We further show that this algorithm consistently retrieves the missing data even during massive data loss situations. 
%As we deploy more wireless sensors in various application domains, the probability of generating missing data also increases. This incomplete data could lead to catastrophic consequences if used for decision-making. There is rich literature dealing with this problem. However, most approaches show performance degradation when there is sizable missing data. Inspired by the emerging field of graph signal processing, this paper studies a new Sobolev method to recover missing sensor data based on the reconstruction of time-varying graph signals. The proposed method retrieves the lost data by considering the spatial and temporal relationships in the available data. Experimental comparisons on several publicly available datasets show that the  Sobolev algorithm outperforms a number of previous works by a maximum margin of 54\%. We further show that the algorithm consistently recovers the missing data even during massive data loss situations. 
\vspace{1mm} 

\end{abstract}

\begin{IEEEkeywords}
Sensor Networks, Graph Signal Processing, Time-varying Graph Signal Reconstruction, Missing Data Recovery
\vspace{-2mm}
\end{IEEEkeywords}

\section{Introduction}

Wireless sensor networks (WSNs) are networks of linked sensor nodes that interact wirelessly to collect information of interest. These wireless nodes are tiny battery-powered devices that consume low power and are easy to deploy. For these reasons, we widely use WSNs for object detection, weather monitoring, pollution monitoring, security surveillance \cite{b3} and many other tasks. However, because of low cost and use in remote locations, often it becomes challenging to repair the malfunctioning nodes. Also, since the number of nodes often crosses the range of thousands \cite{b3}, manually replacing or repairing faulty sensors becomes a tedious task. Since missing data severely hamper the process of decision-making, it is necessary to find out those missing values.

There are several approaches in the literature to deal with this problem. One of the widely used methods is k-nearest neighbors interpolation (kNN) \cite{b13, b25}. kNN recovers the missing data by weighing and averaging the readings of the nearest k neighboring nodes. Some statistical methods based on Expectation-Maximization (EM) algorithm\cite{b16} are also proposed. Here the missing data is recovered by alternately computing the conditional expectation in E-step and updating the estimate by maximizing this conditional expectation in the M-step. Even though these approaches are easy to implement and have beaten several contemporary state-of-the-art methods, they fail to consider the spatio-temporal relationship in data, thereby restricting their implementation in real-life situations. As an improvement, few studies have approached the spatial-temporal data recovery problem as a low-rank matrix completion (LRMC) \cite{b18, b19} problem. Here they estimate the missing values by finding an appropriate low-rank approximation of the original incomplete data matrix.
Nonetheless, these methods only capture the superficial features, failing to infer the embedded spatio-temporal dependencies in the sensor data. In the recent literature, some promising data-recovery methods are proposed, like the Probabilistic Matrix Factorization (PMF) \cite{b27}. In this method, the sensors are divided into different groups based on their similarity using k-means clustering, and then a PMF algorithm is applied within each group to recover the missing data. Even though this algorithm gives promising results, the performance of this algorithm is severely dependent upon the choice of hyperparameter k in k-means clustering and is prone to overfitting \cite{b2}.

In recent years, due to the increasing demands for signal and information processing in irregular domains, the use of graph-based methods has increased in various fields, \textit{e.g.} computer vision \cite{b10, b12}, biological networks, point cloud processing, data science\cite{b11}, prediction of infectious diseases \cite{b1} and others. Sensors network is one such domain where graph signal processing (GSP) finds its most natural applications \cite{b11, b24}. GSP tries to extend the concept of classical digital signal processing to graphs. When GSP is applied in sensor networks, the graph vertices represent the position of sensor nodes, whereas the sensor attributes (readings like temperature, pressure, and humidity) are represented by a vertex-indexed signal, known as graph signals \cite{b24}. In this paper, we show that GSP-based methods can be useful to interpolate missing sensor data. Here we formulate the problem of missing sensor data as a reconstruction of time-varying graph signals \cite{b1, b4}. Our method is built upon the work of Giraldo \textit{et al.} \cite{b1}, which predicts the number of new COVID-19 cases by extending the Sobolev norm \cite{b21} defined in GSP for time-varying graph signals \cite{b1}. The Sobolev algorithm successfully interpolates missing sensor values even in situations involving massive data loss. Our method also significantly improves upon the baselines while tested on several publicly available datasets. Our contributions can be summarized as follows:
\begin{itemize}
\item We introduce the concepts  of  reconstruction  of graph  signals  from  GSP for recovering missing data in wireless sensor networks. 
\item The Sobolev algorithm shows significant performance improvement over state-of-the-art (SoTA) algorithms dealing with the problem of missing sensor data.
\item We test the studied method on several publicly available datasets from diverse environments (indoor and outdoor) to demonstrate its versatility.  
\end{itemize}

We organize the rest of the paper as follows: In Section II, we discuss GSP basics and then gradually move into the details of the studied method. The experimental framework (including datasets, evaluation metrics, experiments, and results) is briefly discussed in Section III. Finally, in Section IV, we conclude the paper with outlines for the direction of future work.

\vspace{-1mm}
\begin{figure*}[!h]
    \centering
    \includegraphics[width =0.9\linewidth]{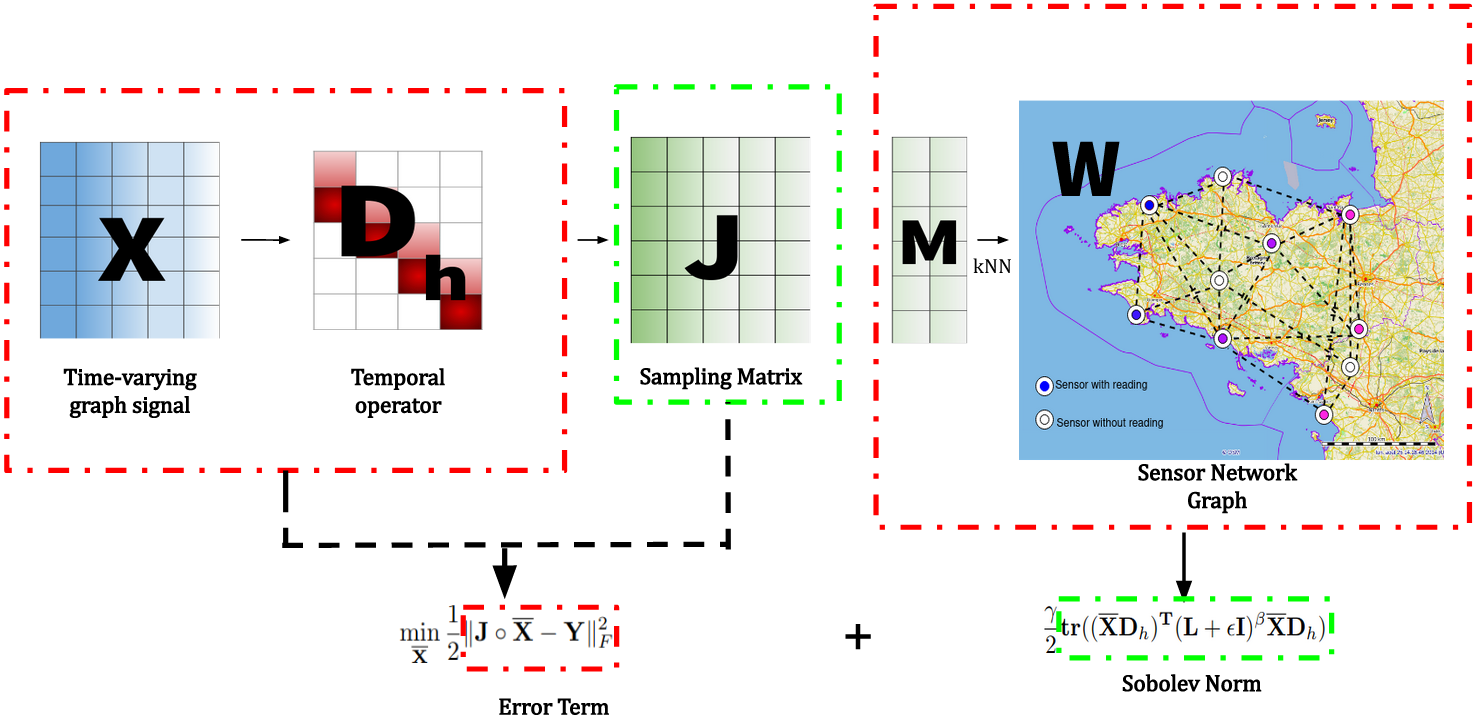}
    
    \caption{Schematic diagram of the proposed method.}
    \label{fig: schematic}
\vspace{-3mm}
\end{figure*}

\section{Preliminaries and Proposed Work}
This section presents the mathematical notations used in this paper, along with the basics of graph signal processing. Further, we discuss the framework to recover missing data in sensor networks which is inspired by the theory of reconstruction of time-varying graph signals \cite{b4, b14} by minimizing the Sobolev norm \cite{b1}.
\vspace{-1mm}
\subsection{Notation}
For the convenience of readers, firstly we introduce the various notations used in this paper. The uppercase boldface letters like $\mathbf{X}$ represents the matrices, lowercase boldface letters like $\mathbf{x}$ denote vectors. Calligraphic letters like $\mathcal{L}$ denotes sets. Matrix products like Hadamard and Kronecker products are denoted by $\circ$ and $\otimes$, respectively. The vectorization of \textbf{X} is denoted by vec(\textbf{X}). $(\textbf{X})^T$ denotes the transpose of matrix \textbf{X}. The diagonal of a matrix with entries $ x_1, x_2, x_3,....,x_n$ is denoted by diag(\textbf{x}). The $\mathit{\ell}_2$ and Frobenius norm of a vector \textbf{x} are represented by $\lVert \textbf{x} \rVert_2$ and $\lVert \textbf{x} \rVert_F$ respectively. The trace of a matrix \textbf{X} is given by $tr(\textbf{X})$. Any other mathematical symbol used in this paper bear their usual significance. 

\vspace{-1mm}
\subsection{Graph Construction and Graph Signals Representation}\label{AA}
Let us consider an undirected weighted graph $G = (\nu, \mathcal{E}, \textbf{W})$. Here $\nu$ is the set of $N$ nodes/ vertices with $|\nu| =N$. $\mathcal{E} = \{(i,j)\}$ represents the set of edges with $(i,j)$ as an edge between the nodes $i$ and $j$. $\textbf{W}$ is the weighted adjacency matrix.

According to the spectral graph theory \cite{b7} and graph signal processing \cite{b11}, we define the unnormalized Laplacian of $G$ as $\mathbf{L} = \mathbf{D} - \mathbf{W}$, where $\mathbf{D}$ is the diagonal matrix of $G$ and $\mathbf{D}(i,i) = \sum_{j=1}^{N} \mathbf{W}(i,j)$ $\forall i = 1,....,N$. The eigenvalues of $\mathbf{L}$ are $0 = \lambda_1 \leq \lambda_2 \leq .... \leq \lambda_N $ and their corresponding eigenvectors are $\{\textbf{u}_1, \textbf{u}_2, .... , \textbf{u}_N\}$.  

Finally, on each node, we define a graph signal that is a function on $\nu$ such that $x: \nu \rightarrow \mathbb{R} $. The graph signal can be represented as a vector $\mathbf{x} \in \mathbb{R}^N$, where $\mathbf{x}(i)$ is the value of the function in the node $i \in \nu$.

\subsection{Reconstructing time-varying graph signals by Sobolev Norm Minimization}

In the theory of graph signal processing, sampling and reconstructing graph signals play an essential role \cite{b11}. For reconstructing a graph signal from its samples, the graph signals need= [ to be bandlimited \cite{b9}. In the literature of GSP, most of the proposed recovery algorithms make a prior assumption that graph signals are smooth in graphs. In our work involving sensor networks, we also consider graph signals to be spatiotemporally smooth.
%In the context of sensor networks, the sensor readings, which are the graph signals, are usually smooth on graphs,  rather than being strictly bandlimited. 
This is because adjacent sensor readings are similar and the readings are progressive with time. For static graph signals in $G$, the graph Laplacian quadratic form $S_2(\mathbf{x})=\mathbf{x}^T\mathbf{Lx}$ is widely used as a measure of smoothness\cite{b8}. However, in sensor networks, missing data may arise at any point in space-time; hence we need to consider both the spatial and the temporal distribution of missing data while we try to reconstruct their original values. So the concept of time-varying graph signals is ideal for dealing with these forms of missing data. Qiu \textit{et al.} \cite{b4} extended the definition of $S_2(\textbf{x})$ to time-varying graph signals. Let a time-varying graph signal be expressed as a matrix ${\mathbf X} =[{\mathbf x}_1,{\mathbf x}_2,\ldots, {\mathbf x}_M]^T$, where $\mathbf{x}_t$ denotes the signal at time $t$ $(1 < t < M)$ in $G$. Here each row of $\mathbf{X}$ represents a time-series on the corresponding vertex. Now we define the smoothness of $\mathbf{X}$ as \cite{b4}:
\begin{equation} S_2({\mathbf X}) = \sum \limits _{t=1}^{M}S_2({\mathbf x}_t) = {\rm tr}\left({\mathbf X}^{\rm T}{\mathbf L}{\mathbf X}\right). \label{eq: smoothness} \end{equation}

As per Qiu \textit{et al.}, the temporal difference operator matrix $\mathbf{D}_{h}\in \mathbb{R}^{M\times(M-1)}$ is defined as:

\begin{equation} \mathbf{D}_{h}=\begin{bmatrix} -1 & & & \\ 1 & -1 & & \\ & 1 & \ddots & \\ & & \ddots & -1 \\ & & & 1 \end{bmatrix} \in \mathbb{R}^{M \times(M-1)}. \label{eq: tempdefmatrix}  \end{equation}

This is done so as to include temporal information in the problem of time-varying graph signal reconstruction. Also, the temporal difference signal is defined as:

\begin{equation} \mathbf{XD}_{h}=[\mathbf{x}_{2}-\mathbf{x}_{1},\ \mathbf{x}_{3}-\mathbf{x}_{2},\ \ldots,\ \mathbf{x}_{M}-\mathbf{x}_{M-1}]. \label{eq: tempdefsignal} \end{equation}

In their work, Qiu \textit{et al.} had proposed two different reconstruction methods, one for the noisy case and the other for the noiseless case. Since sensor networks accumulate sufficient noise during their operational period, we consider the noisy case in our work. As per Qiu \textit{et al.}, the noisy case is defined as follows:
%\vspace{-1.5mm}
\begin{equation} \min_{{\overline{\mathbf{X}}}}\frac{1}{2}\Vert \mathbf{J}\circ\overline{\mathbf{X}}-\mathbf{Y}\Vert_{F}^{2}+\frac{\gamma}{2}\mathbf{tr} ((\overline{\mathbf{X}}\mathbf{D}_{h})^{T}\mathbf{L}\overline{\mathbf{X}}\mathbf{D}_{h}), \label{eq: qiu} \end{equation}

where $\mathbf{Y}$ is the sampled matrix. The sampling matrix for the whole time-varying graph signal is $\mathbf{J}\in\{0,1\}^{N\times M}$, defined as:
%\vspace{-1.5mm}
\begin{equation} \mathbf{J}(i,\ j)=\begin{cases} 1\ \mathbf{if}\ i\in S_{t},\\ 0\ \mathbf{if}\ i\not\in S_{t}, \end{cases} \label{eq: samplingmatrix} \end{equation}

${S_t}$ being the sampled set of vertices at time \textit{t}. In a nutshell, Eqn. \ref{eq: qiu} reconstructs a time-varying graph signal $\overline{\mathbf{X}}$  with a small error $\Vert \mathbf{J\circ \overline{X}-Y}\Vert_{F}^{2}$ while minimizing the temporal difference graph signal smoothness $tr(\overline{\mathbf{X}}\mathbf{D}_{h})^{T}\mathbf{L}\overline{\mathbf{X}}\mathbf{D}_{h})$. $\gamma$ in Eqn. \ref{eq: qiu} is known as the regularization parameter and it weights the importance between the error and smoothness terms. 

In 2020, Giraldo and Bouwmans \cite{b1} proposed a new algorithm for time-varying graph signals reconstruction inspired by the minimization of the Sobolev norm. The norm was first defined by Pesenson \textit{et al.} \cite{b21} for introducing the variational problem in graphs as: 

\textbf{Sobolev norm}: Let $\mathbf{L}$ and $\mathbf{x}$ be the Laplacian matrix and graph signal, respectively. Let $\epsilon \geq 0$ and $\beta \in \mathbb{R}^+$ be two constant parameters, and the Sobolev norm is defined as follows:
    \begin{equation} \Vert \mathbf{x}\Vert_{\beta,\epsilon}=\Vert(\mathbf{L}+\epsilon \mathbf{I})^{\beta/2}\mathbf{x}\Vert. \label{eq: sobolev} \end{equation}
    When $\mathbf{L}$ is symmetric, we have that as:
    \begin{equation} \mathbf{x}^{\mathbf{T}}(\mathbf{L}+\epsilon \mathbf{I})^{\beta}\mathbf{x}. \label{eq: symmetric} \end{equation}

In their work, they found that the term $(\mathbf{L}+ \epsilon \textbf{I})$ in Eqn. \ref{eq: symmetric} has a better condition number than $\mathbf{L}$ when $\epsilon>0$. By extending the concept of Sobolev-norm to time-varying graph signals (considering $\mathbf{L}$ to be a symmetric matrix), we get:
\begin{equation} \Vert \mathbf{X}\Vert_{\beta,\epsilon}=\sum_{i=1}^{M}\mathbf{x}_{i}^{\mathbf{T}}(\mathbf{L}+\epsilon \mathbf{I})^{\beta}\mathbf{x}_{\mathbf{i}}=\mathbf{tr} (\mathbf{X}^{\mathbf{T}}(\mathbf{L}+\epsilon \mathbf{I})^{\beta}\mathbf{X}). \label{eq: sobolevtv} \end{equation}

Finally, the Sobolev reconstruction problem for time-varying graph signals is formulated as:

\begin{equation} \min_{\overline{\mathbf{X}}}\frac{1}{2}\Vert \mathbf{J}\circ\overline{\mathbf{X}}-\mathbf{Y}\Vert_{F}^{2}+\frac{\gamma}{2}\mathbf{tr} ((\overline{\mathbf{X}}\mathbf{D}_{h})^{\mathbf{T}}(\mathbf{L}+\epsilon \mathbf{I})^{\beta}\overline{\mathbf{X}}\mathbf{D}_{h}). \label{eq: sobolevrectv} \end{equation}

Here we use the temporal difference operator of Eqn. \ref{eq: tempdefmatrix} along with the Sobolev-norm of time-varying graph signals of Eqn. \ref{eq: sobolevtv}. The schematic diagram of our method is shown in Fig. \ref{fig: schematic}.

\begin{table}
\centering
\caption{Performance of the proposed method under different sampling densities in the two datasets}
\label{tab:sampdensity vs Dataset}
\makebox[\linewidth]{
\scalebox{1}{
\begin{tabular}{|c|c|c|c|c|} 
\hline
\multirow{2}{*}{Sampling Density} & \multicolumn{2}{c|}{Mol\'ene Dataset} & \multicolumn{2}{c|}{Intel Dataset} \\ 
\cline{2-5}
 & \textbf{RMSE} & \textbf{MAE} & \textbf{RMSE} & \textbf{MAE} \\ 
\hline
0.1 & 2.40 & 1.65 & 3.04 & 1.50 \\ 
\hline
0.3 & 1.60 & 1.02 & 2.01 & 0.97 \\ 
\hline
0.5 & 1.20 & 0.74 & 1.57 & 0.70 \\ 
\hline
0.7 & 0.97 & 0.65 & 1.30 & 0.55 \\
\hline
\end{tabular}
}
}
\vspace{-4mm}
\end{table}

\begin{table*}
\centering
\caption{Quantitative comparison results of SoTA methods vs proposed method. Best results are in \textbf{bold}.}
\label{tab:metric table}
\makebox[2\columnwidth]{
\scalebox{1.15}{
\label{tab:metric table}
\begin{tabular}{|c|c|c|c|c|c|c|c|c|c|c|c|} 
\hline
\multirow{2}{*}{\begin{tabular}[c]{@{}c@{}}\\\end{tabular}} & \multirow{2}{*}{\begin{tabular}[c]{@{}c@{}}Sampling \\Density\end{tabular}} & \multicolumn{2}{c|}{kNN \cite{b13, b25}} & \multicolumn{2}{c|}{EM \cite{b16}} & \multicolumn{2}{c|}{LRMC \cite{b18, b19} } & \multicolumn{2}{c|}{PMF \cite{b27}} & \multicolumn{2}{c|}{Proposed}      \\ 
\cline{3-12}
                                                            &                                                                             & RMSE & MAE               & RMSE & MAE                & RMSE & MAE                      & RMSE & MAE               & RMSE          & MAE            \\ 
\hline
                                                            & 0.1                                                                         & 8.05 & 6.75              & 7.34 & 5.68               & 3.47 & 2.35                     & 4.75 & 2.56              & \textbf{2.40} & \textbf{1.65}  \\ 
\cline{2-12}
Molene Dataset                                              & 0.3                                                                         & 5.45 & 4.68              & 5.20 & 3.93               & 2.82 & 1.35                     & 3.42 & 1.64              & \textbf{1.60} & \textbf{1.02}  \\ 
\cline{2-12}
                                                            & 0.5                                                                         & 4.20 & 3.04              & 4.14 & 2.95               & 1.96 & 1.02                     & 2.45 & 1.09              & \textbf{1.20} & \textbf{0.74}  \\ 
\cline{2-12}
                                                            & 0.7                                                                         & 3.40 & 2.13              & 3.19 & 1.82               & 1.02 & 0.81                     & 1.22 & 0.95              & \textbf{0.97} & \textbf{0.65}  \\ 
\hline
                                                            & 0.1                                                                         & 9.70 & 5.34              & 7.78 & 5.63               & 4.68 & 2.94                     & 5.34 & 3.98              & \textbf{3.04} & \textbf{1.50}  \\ 
\cline{2-12}
Intel Dataset                                               & 0.3                                                                         & 6.30 & 3.52              & 6.45 & 4.96               & 3.40 & 2.15                     & 3.95 & 2.37              & \textbf{2.01} & \textbf{0.97}  \\ 
\cline{2-12}
                                                            & 0.5                                                                         & 4.60 & 2.98              & 5.17 & 3.81               & 2.06 & 1.21                     & 2.75 & 1.65              & \textbf{1.57} & \textbf{0.70}  \\ 
\cline{2-12}
                                                            & 0.7                                                                         & 3.75 & 1.78              & 4.63 & 2.57               & 1.45 & 0.75                     & 1.97 & 0.99              & \textbf{1.30} & \textbf{0.55}  \\
\hline
\end{tabular}
}
}
\vspace{-4.75mm}
\end{table*}

\section{Experimental Framework}

This section briefly discusses the implementation of our proposed method and the dataset used to assess it. We also discuss the evaluation metrics and compare our method with SoTA baselines \cite{b13, b25, b16, b18, b19, b27} on those metrics.

All the experiments were performed in MATLAB\textsuperscript{\tiny\textregistered} 2018a on a Intel\textsuperscript{\tiny\textregistered} Core\textsuperscript{TM} i5 8th Gen. Laptop with Linux Mint and having 4 CPUs with a clock frequency of $1.6$ GHz each. We used Python 3.10.0 to generate the figures and graphs in the paper.

\subsection{Datasets}

To evaluate the performance of our proposed method, we test it on many publicly available widely used sensor-network datasets. To test the flexibility of the proposed method, we use datasets collected in diverse environments (indoor and outdoor). 

\textbf{Mol\'ene Dataset:} \cite{b5} The French national meteorological service\footnote{\url{http://data.gouv.fr}} published an open-access dataset of hourly weather observations in Brittany, France, for the month of January 2014. In addition to the graph of ground weather stations, the dataset contains hourly readings of those stations. Readings include temperatures, wind characteristics, rain, and other information. 

\textbf{Intel Lab Dataset:} \cite{b23} This dataset contains data collected from 54 sensors deployed in the Intel Berkeley Research lab between February 28th and April 5th, 2004\footnote{\url{http://db.csail.mit.edu/labdata/labdata.html}}. Sensors measured various parameters like temperature, humidity, light intensity, and voltage values at an interval of 30 seconds in an indoor environment. For convenience in processing and representing the data, we consider the temperature readings only.  

\subsection{Evaluation Metrics}
To compare our method against baselines, we use two of the most widely used metrics in sensor networks, namely the \textbf{Root Mean Square Error (RMSE)} and the \textbf{Mean Absolute Error (MAE)}. The two metrics are defined as:

\begin{equation}
    RMSE = \sqrt{\frac{1}{n}\sum_{i=1}^n(x_i - x^*_i)^2}, 
    \label{eq: rmse}
\end{equation}

\begin{equation}
    MAE = \frac{1}{n}\sum_{i=1}^n\vert x_i - x^*_i\vert.
\end{equation}

Here $n$ is the number of unavailable/ missing readings, $x_i$ is the ground truth (original) data, and $x^*_i$ is the reconstructed/ recovered data using the proposed method.

\subsection{Experiments and Results}
Let $\mathbf{M}\in \mathbb{R}^{N\times 2}$ be the matrix of positions of all sensor nodes in $\nu$ such that $\mathbf{M}=\lceil \mathbf{m}_{1}, \ldots, \mathbf{m}_{N}\rceil^{\mathit{T}}$, where $\mathbf{m}_{i}\in \mathbb{R}^{2}$ is the vector with the latitude and longitude (for the Mol\'ene dataset) or $x$ and $y$ coordinates (for the Intel dataset) of vertex $i$. To connect the vertices of the graph, we use a k-nearest neighbors (kNN) strategy. The value of $k$ is determined experimentally. The weights of each edge $(i,j)$ connecting the nodes $i$ and $j$ is given by $\textbf{W}(i,j)=\exp(-\frac{\delta(i,j)^{2}}{\sigma^{2}})$, where $\delta(i,j)=\Vert \mathbf{m}_{i}-\mathbf{m}_{j}\Vert_{2}$ is the Euclidean distance between the nodes $i$ and $j$ and $\sigma^2$ is the standard deviation of the Gaussian kernel $\sigma=\frac{1}{\vert \mathcal{E}\vert}\sum_{(i,j)\in \mathcal{E}}\delta(i,j)$.
\vspace{1.1mm}

We randomly remove some data points from the dataset and then recover them for evaluating the Sobolev reconstruction model's performance. For a fair comparison, we test the baseline methods on the same datasets on which we test our approach. We further scale the data to the range $[0,1]$. For the sampling matrix $\mathbf{J}$, we take a random sampling strategy so that each time-graph signal $\mathbf{x}_t$ has the same number of sampled nodes $\forall t$ ($1<t<M$). Also, we empirically tune the values of constants $\epsilon$, $\gamma$, and $\beta$ to get the combination of values for which the performance of our proposed method is the best. Again, for fairly comparing the studied algorithm with the baselines (kNN, Expectation-Maximization, LRMC, and PMF), we find out the best values of different parameters affecting their performances. 

\begin{figure*}[!h]
    \centering
    \includegraphics[width = \linewidth]{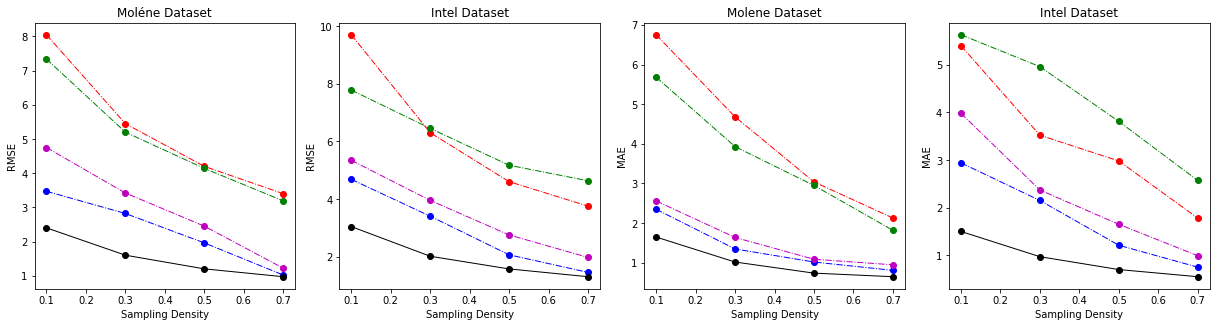}
    \includegraphics[width = 0.67\textwidth]{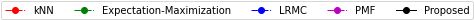}
    
    \caption{Performance of various SoTA algorithms in recovering missing sensor data.}
    \label{fig:final graph}
\vspace{-4.75mm}
\end{figure*}

For the Mol\'ene dataset, the value of nearest neighbors is determined experimentally and is set to $k = 5$. Since all the weather stations do not have readings over the whole period of $31$ days, we consider the temperature readings of such stations where the readings are consistently available. From the $744$ hours of readings, we randomly remove some readings while using the available data from the rest of the dataset. The sampling densities are set to $\{0.1, 0.3, 0.5, 0.7\}$, where each sampling density means the amount of valid information. For example, when the sampling density is $0.1$ the reconstructed data is $90 \%$, and the available data is $10 \%$. 

For the Intel dataset, the value of $k$ is set to $3$. Here we consider the first $10^4$ readings out of $10^5$ readings ($38$ days $\times$ $24$ hours $\times$ $60$ minutes $\times$ $2$). In this dataset too, we randomly remove some values with sampling densities $\{0.1, 0.3, 0.5, 0.7\}$ and reconstruct those values using our method.

For both datasets, we perform Monte-Carlo cross-validation \cite{b15} with 20 repetitions. Finally we compare the reconstructed values with the ground truth data and evaluate the performance of our algorithm.

Table \ref{tab:sampdensity vs Dataset} shows the performance of our algorithm under different sampling densities. Here we observe that our method performs well under higher sampling densities than under lower ones. Since the number of available readings increases with increasing sampling density, we are getting richer information about the underlying spatio-temporal information in the dataset. So, during recovering the missing values, we are getting a better performance. In Table \ref{tab:metric table}, we compare the proposed method against the baselines \cite{b13, b25, b16, b18, b19, b27} with different sampling densities. Fig \ref{fig:final graph} shows that even though other methods fail to recover the missing values under lower sampling densities accurately, our method significantly outperforms them by providing consistent performance under those conditions. One explanation behind our model's success can be its ability to capture the spatial and temporal relation in the data effectively. Moreover, there is no prior knowledge about the behavior of the underlying  processes in the time domain affecting the sensor readings. Even then, our method successfully recovers the missing values.

\section{Conclusion and Future Work}
In this paper, we introduce a time-varying graph signals reconstruction approach to solve the problem of missing sensor data recovery in wireless sensor networks. We show how the proposed method performs well under massive data loss situations, even without prior knowledge of the factors affecting the sensor readings. We also validate the proposed method on several publicly available datasets belonging to diverse environmental conditions to demonstrate its flexibility. We believe that our method will help future researchers to use the promising field of GSP in solving various related problems like intelligent transportation systems, weather conditions forecasting, disaster management, and others.

We plan to extend the studied algorithm to non-smooth situations, for example, where nearby localities have dissimilar readings (e.g., in rugged terrain) and changes in weather conditions are abrupt with time. Using graph learning can also be a way forward. 

\section{Acknowledgement}
Anindya Mondal would like to thank Jhony H. Giraldo of MIA Lab, La Rochelle Universit\'e, France, who proposed the Sobolev norm minimization method for reconstructing time-varying graph signals \cite{b1}, for sharing his valuable insights in various sections of this paper. We also thank the Texas Instruments Innovation Laboratory, Jadavpur University, for the resources used to carry out the experiments.

\vspace{12pt}
%\color{red}
%IEEE conference templates contain guidance text for composing and formatting conference papers. Please ensure that all template text is removed from your conference paper prior to submission to the conference. Failure to remove the template text from your paper may result in your paper not being published.


\begin{thebibliography}{00}

\bibitem{b3} Chatterjee A, Venkateswaran P, Das K. Simultaneous State Estimation of Cluster-Based Wireless Sensor Networks. IEEE Transactions on Wireless Communications, 2016 
\bibitem{b13} K. Kiani and K. Saleem, "K-nearest Temperature Trends: A Method for Weather Temperature Data Imputation," in Proc. Int. Conf. Inf. Syst. Data Mining (ICISDM), 2017
\bibitem{b25} Rukundo O, Cao H. Nearest Neighbor Value Interpolation. arXiv preprint arXiv:1211.1768. 2012
\bibitem{b16} K. Zhang, R. Gonzalez, B. Huang, and G. Ji, "Expectation – Maximization Approach To Fault Diagnosis with Missing Data," IEEE Trans. on Industrial Electronics, 2015
\bibitem{b18} M. T. Asif, N. Mitrovic, J. Dauwels, and P. Jaillet, "Matrix and Tensor Based Methods for Missing Data Estimation in Large Traffic networks," IEEE Trans. Intell. Transp. Syst., 2016
\bibitem{b19} R. López-Valcarce and J. Sala-Alvarez, "Low-Rank Data Matrix Recovery With Missing Values And Faulty Sensors," 27th EUSIPCO, 2019
\bibitem{b27} Fekade B, Maksymyuk T, Kyryk M, Jo M., "Probabilistic Recovery of Incomplete Sensed Data in IoT", IEEE Internet of Things Journal, 2017
\bibitem{b2} Du, J., M. Hu, and W. Zhang, "Missing Data Problem in the Monitoring System: A Review", IEEE Sensors Journal, 2020
\bibitem{b10} Mondal A, Shashant R, Giraldo JH, Bouwmans T, Chowdhury AS. Moving Object Detection for Event-based Vision using Graph Spectral Clustering. In Proc. of the IEEE/CVF ICCV 2021
\bibitem{b12} Mondal, A., and M. Das "Moving Object Detection for Event-based Vision using k-means Clustering." In IEEE 8th UPCON, 2021
\bibitem{b11} A. Ortega, P. Frossard, Jelena K, JMF Moura, and P. Vandergheynst. Graph Signal Processing: Overview, Challenges, and Applications. Proc. of the IEEE, 2018
\bibitem{b1} Giraldo, J. H., and T. Bouwmans, "On the Minimization of Sobolev Norms of Time-varying Graph Signals: Estimation of New Coronavirus Disease 2019 Cases", In 30th IEEE MLSP, 2020
\bibitem{b24} Jabłoński I. Graph Signal Processing in Applications to Sensor Networks, Smart Grids, and Smart Cities. IEEE Sensors Journal, 2017
\bibitem{b4} Qiu, K., X. Mao, X. Shen, X. Wang, T. Li, and Y. Gu, "Time-varying Graph Signal Reconstruction" IEEE Journal of Selected Topics in Signal Processing, 2017
\bibitem{b21} I. Pesenson, "Variational Splines and Paley-Wiener Spaces on Combinatorial Graphs," Constructive Approximation, 2009
%\bibitem{b6} Deng, Y., C. Han, J. Guo, and L. Sun. "Temporal and Spatial Nearest Neighbor Values Based Missing Data Imputation in Wireless Sensor Networks." Sensors, 2021
\bibitem{b14} A. Ortega, Introduction to Graph Signal Processing. Cambridge: CUP, 2021
\bibitem{b7} Chung, Fan RK, and Fan Chung Graham. Spectral Graph Theory. No. 92. American Mathematical Soc., 1997
\bibitem{b9} Chen, S., Varma, R., Sandryhaila, A. and Kovačević, J., Discrete Signal Processing on Graphs: Sampling Theory. IEEE TSP, 2015 
\bibitem{b8} Shuman DI, Narang SK, Frossard P, Ortega A, Vandergheynst P. The Emerging Field of Signal Processing on Graphs: Extending High-dimensional Data Analysis to Networks and Other Irregular Domains. IEEE SPM, 2013
\bibitem{b5} Girault, B., "Stationary Graph Signals Using an Isometric Graph Translation" In 23rd IEEE EUSIPCO, 2015
\bibitem{b23} Madden S. Intel Berkeley research lab data, USA: Intel Corporation, http://db.csail.mit.edu/labdata/labdata.html, 2004
\bibitem{b15} Xu QS, Liang YZ. Monte Carlo cross validation. Chemometrics and Intelligent Laboratory Systems, 2001 
\end{thebibliography}
\end{document}